# Industrial Big Data Analytics: Challenges, Methodologies, and Applications

JunPing Wang, *Member, IEEE,* WenSheng Zhang, YouKang Shi, ShiHui Duan,

*Abstract*—While manufacturers have been generating highly distributed data from various systems, devices and applications, a number of challenges in both data management and data analysis require new approaches to support the big data era. These challenges for industrial big data analytics is real-time analysis and decision-making from massive heterogeneous data sources in manufacturing space. This survey presents new concepts, methodologies, and applications scenarios of industrial big data analytics, which can provide dramatic improvements in velocity and veracity problem solving. We focus on five important methodologies of industrial big data analytics: 1) Highly distributed industrial data ingestion: access and integrate to highly distributed data sources from various systems, devices and applications; 2) Industrial big data repository: cope with sampling biases and heterogeneity, and store different data formats and structures; 3) Large-scale industrial data management: organizes massive heterogeneous data and share large-scale data; 4) Industrial data analytics: track data provenance, from data generation through data preparation; 5) Industrial data governance: ensures data trust, integrity and security. For each phase, we introduce to current research in industries and academia, and discusses challenges and potential solutions. We also examine the typical applications of industrial big data, including smart factory visibility, machine fleet, energy management, proactive maintenance, and just in time supply chain. These discussions aim to understand the value of industrial big data. Lastly, this survey is concluded with a discussion of open problems and future directions.

Note to Practitioners: This paper focus on data acquisition, organization, analyze and decision. Finished industrial big data platform, where main function of massive industrial data ingestion, storage, organization, analyze and decision.

## I. INTRODUCTION

WHEN manufacturers have been entered the age of big data, Data sizes can range from a few dozen terabytes to many petabytes of data in a single data set. For example, the GE company that produces a personal care product generates 5,000 data samples every 33 milliseconds, resulting in [1]. Big data analytics will be vital foundation for forecasting manufacturing, machine fleet, and proactive maintenance. Compared to big data in general, industrial big data has the potential to create value in different sections of manufacturing business chain [2]. For example, valuable information regarding the hidden degradation or inefficiency patterns within machines or manufacturing processes can lead to informed and effective maintenance decisions which can avoid costly failures and unplanned downtime. However, the ability to perform analysis on the data is constrained by the increasingly distributed nature of industrial data sets. Highly distributed data sources bring about challenges in industrial data access, integration, and sharing. Furthermore, massive data produced by different sources are often defined using different representation methods and structural specifications. Bringing those data together becomes a challenge because the data are not properly prepared for data integration and management, and the technical infrastructures lack the appropriate information infrastructure services to support big data analytics if it remains distributed.

Recently, industrial big data analytics has attracted extensive research interests from both academia and industry. According to a report from McKinsey institute, the effective use of industrial big data has the underlying benefits to transform economies, and delivering a new wave of productive growth. Taking advantages of valuable industrial big data analytics will become basic competition for todays enterprises and will create new competitors who are able to attract employees that have the critical skills on industrial big data [3]. The GE Corporate published while book about an industrial big data platform. It illustrates industrial big data requirements that must be addressed in order for industrial operators to achieve the many efficient opportunities in a cost-effective manner. The industrial big data software platform brings these capabilities together in a single technology infrastructure that opens the whole capabilities for service provider [4]. Brian corporate describes industrial big data analytics. The industrial big data analytics will focus on high-performance operational data management system, cloud-based data storage, and hybrid service platforms [5]. ABB corporate proposes that turn industrial big data into decision-making so that enterprise have additional context and insight to enable better decision making [6]. In 2015, industrial big data analytics in [7] is proposed for manufacturing maintenance and service innovation, which discusses automate

J. Wang and W. Zhang are with Laboratory of Precision Sensing and Control Center, Institute of Automation, Chinese Academy, Beijing, China Email: wangjunping@bupt.edu.cn, wensheng.zhang@ia.ac.cn.

S. Duan and Y. Shi are with Communications Standards Research Institute, China Academy of Telecommunication Research of MIIT, Beijing, China Email: duanshihui@ritt.cn, shiyoukang@ritt.cn.

Manuscript received April 30, 2018; revised June, 30 2018.

data processing, health assessment and prognostics in industrial big data environment.

This survey presents new reference model, discusses methodologies challenges, potential solutions and application development of industrial big data analytics. The main contributions are as follows:

1) From a systems-level view, we proposes new industrial big data analytics reference model for manufacturers, which can real-time access, integrate, manage and analyse of massive heterogeneous data sources. The aim of this reference model is to achieve massive predictive manufacturing on three different levels: massive heterogeneous data real-time acquisition and integration, massive industrial data management, intelligent analytics and online decision-making.

2) This survey discusses challenges, development and opportunities of five typical methodologies, which includes Highly distributed industrial data ingestion, large-scale data management techniques, industrial data analytics, industrial big data knowledge repository, and industrial big data governance. 3) This survey also discusses application development and value in four typical area, which include Smart Factory Visibility, Machine Fleet, Energy Management, Proactive Maintenance, Just in Time Supply Chain, and Service Innovation.

The remainder of the survey is organized as follows: Section II will discuss the concept, opportunities and challenges of industrial big data analytics, and presents a reference model and the key challenges of each step in the model. Section III give typical technologies solutions, challenges and development of industrial big data analytics to handle data-intensive applications in Section IV, where categorize the applications of industrial big data analytics into four groups, presenting some application scenarios in each group. Section V finally concludes the article.

## II. THE DATA SOURCES, CHALLENGES AND DEVELOPMENT OF INDUSTRIAL BIG DATA ANALYTICS

### A. Highly Distributed Data Sources

Industrial big data has been produced by diverse sources in manufacturing spaces, such as sensors, devices, logistics vehicles, factory buildings, humans, tacking manufacturing process element(increased production efficiency, factory pollution, reduced energy consumption, and reduced cost of production). Those highly distributed data sources include the following:

*1) Large-scale devices data:* Mobility and the CyberPhysical System(CPS) will change the types of devices that connect into a companies systems and these newly connected devices will produce new types of data. CPS will connect physical items such as sensors, actuators, video cameras and RFID readers to the Internet and to each other. Big data processing and analytics, either on-premise or in the cloud, will collect and analyze data from IoT-enabled devices. These solutions will turn data into context that can be used to help people and machines make more relevant and valuable decisions.

*2) Production life-cycle data:* It includes production requirement, design, manufacturing, testing, sale, maintenance. Products of all kinds of data are recorded, transmitted, and computed, making it possible to product whole life cycle management, to meet the demand of personalized products. Record first, external devices will no longer be the principal means of product data, embedded sensors in the product will get more, real time data products, make product management through requirements, design, production, sales and after-sales to eliminate all useless life course. Second, the interaction between the enterprise and the consumer and trade behavior will produce large amounts of data, mining and analyzing these data, can help consumers to participate in the product demand analysis and product design, flexible manufacturing innovation activities, etc.

*3) Enterprise operation data:* those data includes organization structure, business management, production, devices, marketing, quality control, production, procurement, inventory, goals, plans, e-commerce and other data. Enterprise operation processes data will be innovation enterprise research and development, production, operation, marketing and management style. First of all, production lines and devices data can be used for real-time monitoring of equipment itself, at the same time production data generated by the feedback to the production process, industrial control and management optimization. Second, through the procurement, storage, sales, distribution and following procedures data collection and analysis of supply chain link, will lead to the efficiency of the improved and costs have fallen sharply, and will greatly reduce inventory, improve and optimize the supply chain. Again, the use of the change of the sales data, supplier data, can dynamically adjust the rhythm of the optimal production, inventory and scale. In addition, the energy management system based on real-time perception, can realtime optimize of energy efficiency in the production process.

*4) Manufacturing value chain:* It includes customers, suppliers, partners and other data. To compete in the current global economic environment, enterprises need to fully understand the technology development, production, procurement, sales, services, internal and external logistics competitiveness factors. Big data technology development and application of information flow of each link make up the value chain can be in-depth analysis and mining, for enterprise managers and participants see the new Angle of view of the value chain, makes the enterprise have the opportunity to convert the link of value chain more strategic advantage for the enterprise. Car companies, for example, use large data to predict who will buy specific types of vehicles, so as to realize the target customer response rate increased.

*5) External collaboration data:* It includes economy, industry, market, competitors and other data. In response to

external environmental changes brought about by the risk, the enterprise must fully grasp the current situation of the development of the external environment in order to enhance their ability of strain. Big data analysis technology in the macroeconomic analysis, industry has been more and more widely used in market research, has become the enterprise management decision and an important means of market strain capacity. A handful of leading enterprise has already passed for including from executives to marketing workshop workers even staff to provide information, skills, and tools, guide staff better and more in the "influence" to make decisions in a timely manner.

*B. Challenges of Industrial Big Data Analytics*

Highly distributed data sources bring about challenges in industrial data access, integration, and sharing. Furthermore, massive data produced by different sources are often defined using different representation methods and structural specifications. Bringing those data together becomes key challenge because the data are not properly prepared for data integration and management, and the technical infrastructures lack the appropriate information infrastructure services to support big data analytics if it remains distributed. Those challenges are as follows:

*1) Lack of large-scale data spatio-temporal representation:* In manufacturing fields, every data acquisition device is placed at a specific geographic location and every piece of data has time stamp. The time and space correlation is an important property of data from IoT. During data analysis and processing, time and space are also important dimensions for statistical analysis. The huge volume industrial datasets produced by different sources are often defined using different representation methods and structural specifications. Bringing such data together becomes a challenge because the data are not properly prepared for data spatio-temporal integration and fusion, and the representation technology lack the appropriate information infrastructure services to support analysis of the data if it remains distributed. Statistical inference procedures often require some form of aggregation that can be expensive in distributed architectures, and a major challenge involves finding cheaper approximations for such procedures. Therefore, large-scale data spatio-temporal representing become one challenges for industrial big data analytics.

*2) Lack of both effective and efficient online large-scale machine learning:* The industrial big data generated by industrial IoT has different characteristics compared with general big data because the different types of data collected, of which the most conventional characteristics including heterogeneity, variety, unstructured feature, noise, and high redundancy. Many industrial big data analytics scenarios (e.g., massive detecting machine anomalies and monitoring production quality) require instant answers. Besides just increasing the number of machines to speed up the computation, we need to apply online large-scale machine learning algorithms into a industrial big data analytics framework to provide both an effective and efficient knowledge discovery ability. In addition, traditional data management techniques are usually designed for a single data source. An advanced data management methodology that can organize multiple model data (such as many device status streaming, geospatial, and textual data) well is still missing. Thus, online large-scale machine learning with multiple largescale heterogeneous data becomes one challenges for industrial big data analytics.

*3) Lack of whole processes data life-cycle management:* Cyber-physical systems is generating data at an unprecedented rate and scale that exceed much smaller advances in storage management system technologies. One of the urgent challenges is that the current storage system cannot host the huge data. In general, the value concealed in the industrial big data depends on data timeliness; therefore, we should set up the data quality assurance associated with the analysis value to decide what parts of the data should be archived and what parts should be retired.

*4) Lack of data visualization:* Massive result of industrial big data analytics brings a tremendous amount of information that needs a better presentation. A good visualization of original data could inspire new ideas to solve a problem, while the visualization of analytic results can reveal knowledge intuitively so as to help in decision making. The visualization of data may also suggest the correlation or causality between massive different data factors. The multiple modal in industrial big data analytics scenarios leads to high dimensions of views, such as spatial, temporal, machine and business. The design of such a system is much more challenging than for conventional systems that only reside in one world, as the system needs to communicate with many devices and users simultaneously and send and receive data of different formats and at different frequencies. While there have been advances in visualizing data through various approaches, most notably geographic information system-based capabilities, better methods are required to analyze massive data, particularly data sets that are heterogeneous in nature and may exhibit critical differences in information that are difficult to summarize.

*5) Lack of industrial data confidentiality mechanism:* Most industrial big data service providers could not effectively maintain and analyze such huge datasets because of their limited capacity. They are forced to rely on professionals or tools to analyze such data, which increases the potential safety risks. For example, the transactional datasets generally includes a set of complete operating data to drive key business processes. Such data contains details of the lowest granularity and some sensitive information such as credit card numbers. Therefore, analysis of big data may be delivered to a third party for processing only when proper preventive measures are taken to protect such sensitive data, to ensure its safety.

*C. The Development of Industrial Big Data Analytics*

Above challenges exist in both data management and data analysis that require new industrial big data analytics approaches to support the big data era. Meanwhile these challenges span generation of the data, preparation for analysis, and policy related challenges in its sharing and use. Recently these challenges has attracted extensive research interests from both academia and industry.

Google created GFS [8] and MapReduce [9] models to cope with the challenges brought about by data management and analysis at the internet scale. In addition, massive data generated by users, sensors, and other ubiquitous data sources, which requires a fundamental change on the computing architecture and large-scale data processing mechanism. Luckily industrial companies like Google, Yahoo and Facebook are pushing the envelope on big data needs. Their desire to analyze click streams, web logs, and social interactions has forced them to create new tools for storing and analyzing large data sets. These companies have set up corresponding mechanism which can also be leveraged in the industrial sector to manage the explosion of data that will only continue to grow. Apache Hadoop is one of the most well-established software platforms that support large-scale data management and analytics. This platform consists of the Hadoop kernel, Map/Reduce and Hadoop distributed file system (HDFS), as well as a number of related projects, including Apache Hive, Apache HBase, and so on. Map/Reduce is a programming model and an execution for processing and generating large volume of data sets, is pioneered by Google, and developed by Yahoo and other web companies [9]. Apache Hadoop is one of the most well-established software platforms that support dataintensive distributed applications. It provides a general partitioning mechanism to distribute aggregation workload across different machines. Nevertheless, Hadoop is designed for batch processing.

In industrial manufacturing, Internet of Things has embedded in machines and production line [10]. Prognostics and health management (PHM) is a critical research domain that leverages on advanced predictive tools. large-scale stream processing for real-time analytics is mightily necessary for the PHM [11]. Because industrial stream big data has high volume, high velocity and complex data types, there are a number of different challenges to Map/Reduce framework. Therefore, the industrial real-time big data computing framwork, such as Storm [12], and StreamCloud [13], is very important specially for real-time industrial stream data analytics. Several big data tools based on stream processing have been developed or under developing. One of the most famous platforms is Storm, and others include S4 [14], SQLstream [15], Splunk [16], and SAP Hana [17].

In March 2012, the Obama Administration announced a USD 200 million investment to launch the Big Data Research and Development Plan, which was a second major scientific and technological development initiative after the Information Highway initiative in 1993 [18]. In July 2012, the Vigorous ICT Japan project issued by Japans Ministry of Internal Affairs and Communications indicated that the big data development should be a national strategy and application technologies should be the focus. This paper in [7] discusses on existing trends in the development of industrial big data analytics and CPS, and present that the 5C architecture is necessary steps to fully integrate cyber-physical systems in the manufacturing industry.

Every industrial big data platform has its focus. Some of them are designed for batch processing, some are good at real-time analytic. Each big data platform also has specific functionality, such as statistical analysis, machine learning, and data stream processing. In the above-mentioned Industry 4.0 era, industrial big data analytics and cyber-physical systems are teaming together to realize a new thinking of production management and factory transformation. When all of the data is aggregated, The actual processing of big data into useful information is then the key of sustainable innovation within an Industry 4.0 factory.

## III. TYPICAL TECHNOLOGY OF INDUSTRIAL BIG DATA ANALYTICS

Industrial big data analytics helps us understand the nature of machine states and anomaly even online predict the future of manufacturing. In this section, we discuss five categories of techniques that are frequently used in industrial big data analytics: (1) highly distributed industrial data ingestion, (2) industrial big data repository, (3) large-scale industrial data management, (4) industrial data analytics, (5) industrial data governance.

### A. Industrial Big Data Analytics Reference Model

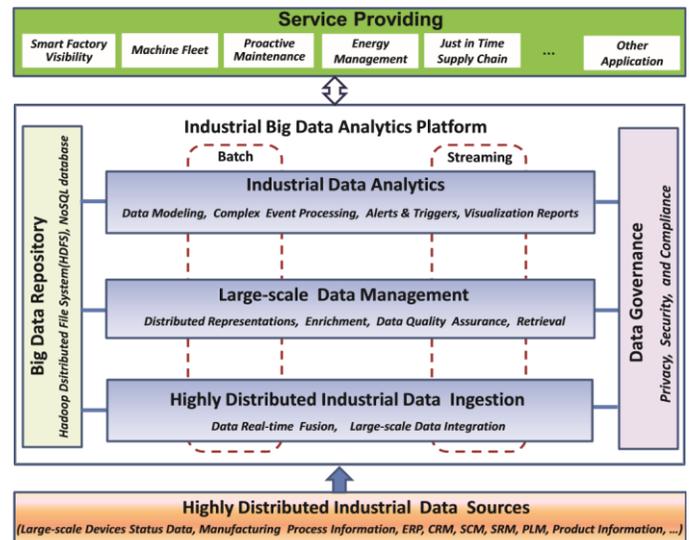

Fig. 1. Reference model of industrial big data analytics, It can be decomposed into three layers, including Highly distributed industrial data ingestion, big data repository, large-scale data management, data governance, industrial data analytics, and service providing, from bottom to up.

Industrial big data analytics reference model is to provide several components of the industrial big data analytics. By segregating layers of responsibilities between different functional components of the model, we can get a clear view of the roles and responsibilities and lay the foundation for a common model of the industrial big data analytics. Fig.1 depicts industrial big data analytics reference model, which consists of several components: massive industrial data ingestion, big data repository, large-scale data management, data governance, industrial data analytics, and service providing. The function of each layer is as follows.

Highly distributed industrial data ingestion is responsible for integrating with a variety of (big) data sources, importing the data into the big data platform and formatting the data into a uniform format. For Big Data, this layer is crucially important to handle the volume, velocity and variety of the data or stream coming into the platform.

Big data repository refers to the storage and management of large-scale datasets while achieving reliability and availability of data accessing. We will review important issues including massive storage systems, distributed storage systems, and big data storage mechanisms.

Large-scale data management can effectively organize largescale heterogeneous data sources, and refers to mechanisms and tools that provide highly efficient knowledge management, such as distributed file systems and SQL or NoSQL data stores.

Data governance encapsulates all other layers of the Big Data Analytics platform to address the best practices. Within some Service Providers, data governance is implemented through a data governance Board. Data governance covers the areas of security, privacy and compliance to legal and regulatory jurisdictions. Data governance defines the policies to be applied to each category of customer or industrial network data and the induced rules to be enforced.

Industrial data analytics implements abstraction application logic and facilitates the data analysis application for service provider. It exploits the interface provided by the programming models to implement various data analysis functions, including querying, statistical analyses, clustering, and classification.

Service providing combines basic analytical methods to develop various filed related applications. It provides five potential big data application domains: devices health care, reducing energy consumed, improve product quality, and custom manufacturing.

*B. Highly Distributed Industrial Data Ingestion*

Manufacturing data ingestion technologies are associated with real-time acquisition and integration of either massive device-generated measurements data (such as status data, performance data, and log file) or enterprise IT software generated data (such as MES, ERP, CRM, SCM, SRM, PLM).

As data volumes increase, the industrial big data analytics platform must allow real-time acquisition and integration of massive heterogeneous data from large-scale industrial devices and software systems. Key technology in manufacturing data ingestion layer is as shows Fig.2:

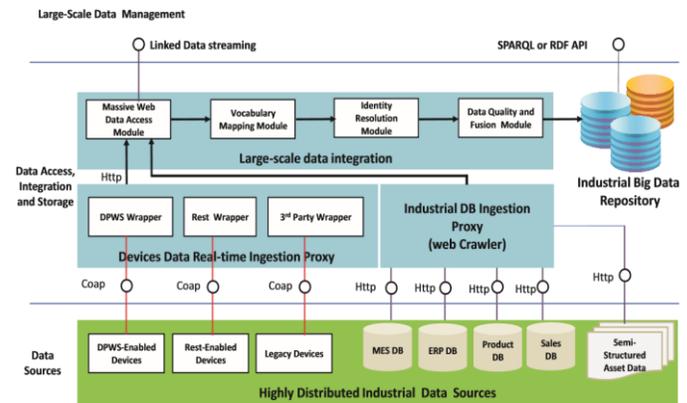

Fig. 2. Highly distributed industrial data real-time acquisition, integration and storage.

In highly distributed manufacturing data acquisition way, On the one hand Real-world service in [19] [20] is embedded into massive industrial asset via SOAP-based web services or RESTful APIs, enabling other components to interact with them dynamically. We can acquire huge heterogeneous data from massive industrial asset, production. On the other hand huge manufacturing processes data files acquisition is accomplished using a combination of web crawler. Web crawler is a program used by search engines for downloading and storing web pages [21]. Generally, web crawler starts from the uniform resource locator (URL) of an initial web page to access other linked web pages, during which it stores and sequences all the retrieved URLs.

Highly distributed manufacturing data integration involved in industrial big data analytics platform include massive web data access module, vocabulary mapping module, identity resolution module and data quality and fusion module. The massive web data access module provides a data crawler as well as components for accessing ontology endpoints and remote RDF dumps. The vocabulary mapping module provides an expressive mapping language for translating data from the various vocabularies that are used on the Web to a consistent, local target vocabulary. The identity resolution component which discovers URI aliases in the input data and replaces them with a single target URI based on flexible, user-provided matching heuristics. The data quality and fusion module which allow web data to be filtered according to different data quality assessment policies and provide for fusing web data using different conflict resolution methods.

Previously, there are two prevailed approaches of data acquisition and integration, which includes the data warehouse method and the data federation method [22]. However, largescale real-time analysis systems not only acquire a data stream from many sources, they also typically collect many distributed data streams and correlate their results. Large-scale data integration framework in [23] can be used to ingest largescale linked data from massive web and to translate the gathered data into a clean local target representation while keeping track of data provenance. The data integration framework also provides a reliable probabilistic mapping approach to entity linking and instance matching [24], which exploits the tradeoff between large-scale automatic instance matching and high quality human annotation. One of these challenges is to rate and to integrate data based on their quality for the system. A new integration of Data Fusion Methodology was done recently by Liu et al [25]. The method develops a systematic data-level fusion methodology that combines the degradation based signals from multiple sensors to construct a health index for better characterizing the condition of a unit. The disadvantage of these method is needed to investigate the performance when creating the nonlinear mappings between the health index and each original sensor data.

*C. Industrial Big Data Repository*

This module provides storage of all data within the big data platform which can be either in the original raw form in which it was ingested into the system or in any intermediate, processed form produced by any other of the reference model layers. The big data repository interacts with all other layers and can be thought as the equivalent to a data bus. In this section, we present a Hadoop-based industrial big data repository, as illustrated in Fig.4.

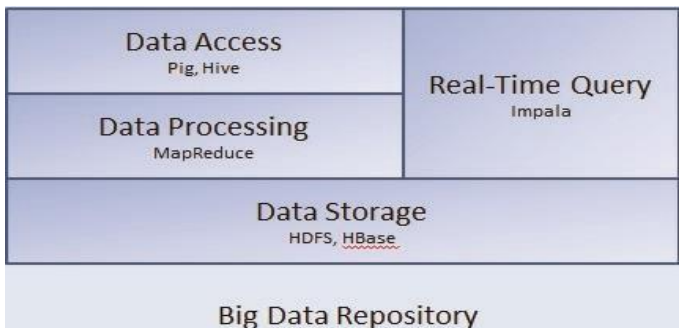

Fig. 3. Hadoop-based industrial big data repository system. It can be organized into four different architectures, including data access, data processing, realtime query and data storage.

The file system is the basis of industrial big data storage and therefore attracts great attention from both industry and academy. In this subsection, we only consider examples that are either open source or designed for enterprise use. The Hadoop Distributed File System (HDFS) provides storage support for the Hadoop Framework. HBase provides additional distributed database functionalities over HDFS. Data stored in HDFS are usually processed with MapReduce operations. Tools like Pig/Hive are developed on top of Hadoop framework to provide data access support over MapReduce/HDFS for upperlevel analytics application. Newer tools like Impala bypasses MapReduce to provide real-time ad-hoc query capabilities. The Data Storage/Processing/Access functionalities provided by the Hadoop ecosystem are table stakes for a Big Data Repository. Database technology has gone through more than three decades of development. Various database systems have been proposed for different scales of datasets and diverse applications. Traditional relational database systems obviously cannot address the variety and scale challenges required by big data. Due to certain essential characteristics, including being schema free, supporting easy replication, possessing a simple API, eventual consistency and supporting a huge amount of data. NoSQL database provide highly available, scalable data storage systems with relaxed consistency guarantees compared to the traditional RDBMS. NoSQL Stores also provide flexible schema to allow heterogeneous columns on different rows of storage.

There are four different types of NoSQL database. The first type is key/value store (e.g. Dynamo, Voldemold), which is inspired by the Amazons Dynamo paper. The data model for this type of database is a collection of key/value pairs. This type of data store provides great flexibility for fast data storage in a programing environment. Second type is column store (e.g. Cassendra, HBase). The data model is based on the original Googles BigTable paper. It stores data tables as columns of data rather than as rows of data. Because of this, it is well suited for Online Analytical Processing (OLAP), which typically involves smaller number of complex queries that access all the data stored. Third type is document store (e.g. MongoDB). It is inspired by the database behind Lotus Notes. Document refers to a collection of information organized in a defined format (XML, JSON, BSON, MS Word etc.). Each document can be retrieved by a key (e.g. a path, a URL). Document store therefore is a collection of key document pairs. Document store allows great flexibility storing semi-structured data and provides query facilities to retrieve documents based on their content. Fourth type of NoSQL store is graph database (e.g. neo4j, Allegro graph). It is inspired by graph theory. Node, a key concept in graph database, is very similar to document in document store. Graph database stores key-node pairs, similar to key-document pairs. In addition, graph database adds relationships between the nodes and also stores key-relationship pairs. Graph database allows graph‑based queries such as shortest path between two nodes and diameter of the graph.

Although MapReduce framework is highly scalable for Big Data queries, it usually does not provide real-time responses needed interactive queries. Some solutions such as Impala attempt to solve the problem using a real-time ad-hoc SQL query processing engine directly over HDFS, bypassing the MapReduce processing to allow shorter response time. Additional optimizations such as compression can be used to accelerate response time further. In May/June 2013, MapR and Cloudera separately announced a new class of engines based directly/indirectly on Apache Solr/Lucene projects. The search feature of these engines allows text searches on data stored in HDFS, thus lower the technical expertise needed to perform Big Data Analytics. Real-life use cases for search include indexing assistance for semi-structured and unstructured data.

### D. Large-scale Industrial Data Management

This section defines a series of harnessing massive manufacturing datasets techniques that can be applied on datasets ingested into the big data platform, such as distributed representations, knowledge enrichment, knowledge manipulation, knowledge retrieval as well as ensuring data quality and security. Fig.3 describes large-scale manufacturing data management.

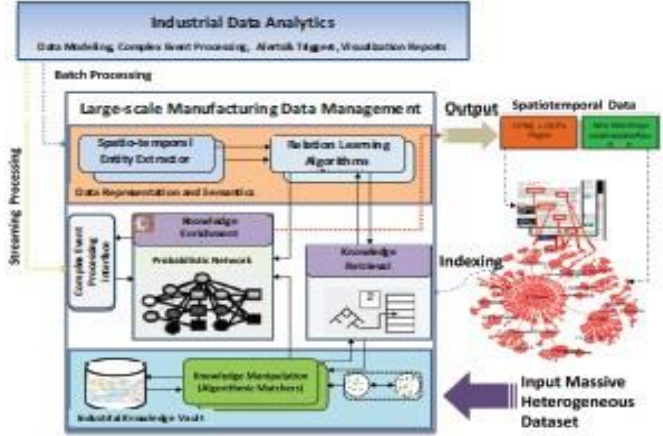

Fig. 4. Large-scale industrial data management system.

The data representation and semantics module map raw data into a manufacturing data model in order to make data meaningful and usable. It can be divided into two parts: spatiotemporal entity extractors and correlation learning algorithms. The spatiotemporal entity extractors extracts spatiotemporal triple from a huge number of data source. Each spatiotemporal entity represents the real manufacturing spcaces entities composed of an identity, descriptive properties and spatial properties. Each extractor assigns a confidence score to an extracted triple, representing uncertainty about the identity of the relation and its corresponding arguments. Recent development of spatio-temporal entity extractors focuses on:(1) active process objects, like for example a production traveling the assembly line of a city, and (2) device status objects, for example, a manufacturing region whose administrative boundary evolves in time. Current main method include Natural Language Processing(NLP) [26], Hierarchical Clustering Network(HCN) [27], lexicosyntactic patterns [28] or lexicosemantic patterns [29], Dynamic Topic Model (DTM) [30], Continuous Time Dynamic Topic Models (cDTM) [31].

The correlation learning algorithms associates different representations and data feature extracted from various sources of the same knowledge entity. For example, this layer can associate the user action knowledge taken from consuming records with the Customer ID taken from the CRM. Both numbers represent the same business entity, customer. Data collected from both sources can be correlated together to provide a richer set of information related to the customer. It also can efficiently estimate the interaction structure from data, such as: traditional clustering algorithms, Gaussian mixtures, nearest-neighbor algorithms, decision trees, or Gaussian SVMs all require $O(N)$ parameters (and/or $O(N)$ examples) to distinguish $O(N)$ input regions. Some typical model has been proposed, like: snapshot model [32], Space Time Composites model (STC) [33] and Spatial-Temporal Object model [34], Event-Based Spatiotemporal Data Model (ESTDM) [35]. However, this model only takes raster data into account, while the causal links between events are hardly picked up in this model. An alternative to ESTDM is the composite processes. The composite process model deals with some of the limitations of the ESTDM. It is designed to represent the links between events and their consequences. Multi-scale graphical models for spatio-temporal processes in [36] is proposed that better represents the multi-scale character of these dynamical systems, efficiently estimate the interaction structure from massive data. Examples of multi-scale spatiotemporal phenomena with the network structure include: flow of information through neural/brain networks [37].

The knowledge enrichment module systematically combines multiple datasets that refer to the same task entity (e.g. customer) in order to create a more complete view of the task entity. We can generate a probabilistic model $p(x,h)$ over the joint space of the latent variables $h$, and observed data or visible variables $x$. Data feature values are conceived as the result of an inference process to determine the probability distribution of the latent variables given the data, i.e. $p(h|x)$, often referred to as the posterior probability. Enrichment learning is conceived in term of estimating a set of model parameters that (locally) maximizes the regularized likelihood of the training data. In some cases, enrichment data sources can be from various customer requirement information databases. In some other cases, some enrichment data can be from the Big Data Analytics results. For example, based on a customers browsing history and locations, it may be inferred with high degree of confidence of the customers gender, age, educational level, income level etc. Main knowledge enrichment methods and standards include semantic enrichment, tensor/matrix factorization [38], Neural Tensor Network [40], The model of TransE [39], Web-scale probabilistic graph [41].

More recently, various types of knowledge enrichment methods have been proposed to embed multi-relational knowledge into low-dimensional enrichment of entities and relations, including tensor/matrix factorization, and neural embedding, bayesian clustering framework. They are hard to explain what relational properties are being captured and to what extent they are captured during the embedding process. Like Knowledge Vault(KV) in [41], KV stores information in the form of RDF triples (subject, predicate, object). Associated with each such triple is a confidence score, representing the probability that the triple is correct. The knowledge enrichment method has two major advantage: (1) these systems extract triples from a huge number of Web sources. Each extractor assigns a confidence score to an extracted triple, representing uncertainty about the identity of the relation and its corresponding arguments. (2) these systems learn the prior probability of each possible triple, based on triples stored in an existing KB. The [41] and [42] use factor graphs to enrich knowledge probabilistic variables and distributions. Note that new enrichment approach is not bound to this representation, they use series of conditional probabilities only or other probabilistic graphical model, and decided to use factor graphs for their illustrative merits. The framework provides knowledge manipulation functionality of reducing statistical complexity and discovering structure that can be applied to globle and local datasets, for example: Union, Intersection, Sorting, Filtering, Compression, Deduplication/Duplication, Group Series functions, and Aggregation functions.

The knowledge retrieval module is a bitmap index that uses Hash functions to conduct lossy compression storage of data. It has high space efficiency and high query speed, but it also has some disadvantages in recognition and deletion. Hash Index is always an effective method to reduce the expense of disk reading and writing, and improve insertion, deletion, modification, and query speeds in both traditional relational databases that manage structured data, and other technologies that manage semi-structured and unstructured data. More recently, two types of knowledge index model have been used to speedup the information retrieval process, including Linked Open Data(LOD) index, and junction tree retrieval. The LOD index [43] is a declarative information retrieval engine used to speedup the entity retrieval process. While most LOD data sets provide a public SPARQL interface, they are in practice very cumbersome to use due to the very high latency (from several hundreds of milliseconds to several seconds) and bandwidth consumption they impose. The junction tree model [44] is an important graph index, which a collection of random variables and the edges encode conditional independence relationships among random variables by junction tree structured. The vertices in a junction tree are clusters of vertices from the original graph. An edge in a junction tree connects two clusters. Junction trees are used in many applications to reduce the computational complexity of solving graph related problems.

*E. Industrial Big Data Analytics*

The section acquire multi-view knowledge models from massive manufacturing Data in both batch and streaming modes by supporting functionalities, such as Descriptive/Predictive/ Prescriptive Modeling, Complex Event Processing, Generation of alerts and triggers to actions, Visualization Reports Generation.

*1) Descriptive/Predictive/Prescriptive Modeling:* The section tracks manufacturing data provenance, from data generation through data preparation, and explains the past / predicting the future/recommending next best action by utilizing Machine Learning / Data Mining algorithms, such as: Classification/Regression, Summary statistics, Correlations Analyst, Stratified Sampling, Boosted Decision Trees, Stochastic Gradient Descent, Alternating least squares, Orthogonal matching pursuit, Greedy Forward Selection, Orthogonal Matching Pursuit, Clustering, Pattern Mining, Recommenders / Collaborative Filtering, Statistical analysis, and Descriptive/Predictive analytics. We build machines lifecycle model, flexible manufacturing model, production quality assurance, prognostics and health management, smart factory visibility, energy management, just in time supply chain, etc. Those model help manufacturers understand trends and predict future outcomes, and extracts patterns to provide insight and forecasts.

*2) Complex event processing(CEP):* CEP offers the functionality that makes it possible to implement Big Data Analytics scenarios which require real-time processing. More specifically, CEP controls the processing of streaming data, the correlation of occurring events and the calculation of KPIs on an ongoing basis. Driven by user-supplied business rules, CEP generates alerts or triggers for subsequent actions by external systems. Most Complex Event Processor (CEP) solutions and concepts can be classified into two main categories: A Computation-oriented CEP solution is focused on executing online algorithms as a response to event data entering the system. A simple example is to continuously calculate an average based in data on the inbound events. A Detection-oriented CEP solution is focused on detecting combinations of events called events patterns or situations. A simple example of detecting a situation is to look for a specific sequence of events.

These are sometimes called event processing platforms, complex-event processing (CEP) systems, event stream processing (ESP) systems, or distributed stream computing platforms (DSCPs). For example, WSO2 Complex Event Processor, IBM Streams, Oracle Event Processor, SAP Event Stream Processor, etc. The CEP platform can handle very high data throughput rates, up to millions of events or messages per second. Completely abstracting the entire development and deployment processes, it simplifies the rapid creation of event driven applications for any type of real time business solution, enabling the Enterprise to really immerse itself in next generation real time applications, with times to market of minutes, rather than days or weeks.

*3) Generation of alerts and triggers to actions:* The outcomes produced by Data analysis can trigger alerts and actions. Alerts are mainly destined to humans for further consideration. Triggers are mainly destined to other applications or systems that automatically proceed to the corresponding actions. For example, a network performance monitoring application may use CEP to monitor the alarms from network elements. When the alarm number/severity exceeds certain thresholds, the application will generate a critical alarm to the network operator and trigger the policy changes to reroute network traffic away from the affected subset of the network.

*4) Visualization reports generating:* Visualization Reports can be generated in real-time, on a daily/weekly/monthly basis, or on-demand. They can be used to visualize the Big Data Analytics results. The main objective of visualization reports is to represent knowledge more intuitively and effectively by using different graphs [45]. It makes complex data more accessible, understandable and usable. Users may have particular analytical tasks, such as making comparisons or understanding causality, and the design principle of the graphic follows the task. Recently, there are a lot of visualization tools off the shelf that display efficiently these data reports, for example, IBM Watson , SAP Lumira, Oracle Visual Analytics, Data wrapper, Chart JS, Dygraphs, and Google Charts. However, current Big Data visualization tools mostly have poor performances in functionalities, scalability and response time [46]. What we need to do is rethinked the way we visualize Big Data, a different method from what we adopted before. For example, the history mechanisms for information visualization [47] also are data-intensive and need more efficient approaches. Uncertainty can lead to a great challenge to effective uncertainty aware visualization and arise in any stage of a visual analytics process [48]. New framework for modeling uncertainty and characterizing the evolution of the uncertainty information are highly necessary through analytical processes.

*F. Industrial Big Data Governance*

The Industrial Big Data Governance Layer encapsulates all other layers of the Big Data Analytics platform to address the best practices introduced above and provides those functions: Privacy, Security, and Compliance. Privacy addresses the Customers need for transparency, choice, and control by providing visibility and configuration of privacy preferences and practices. Privacy applies to Personal Identification Information (PII) data. Anonymization techniques can transform PII data into non-PII data. Privacy preservation addresses concerns regarding the disclosure of Customer data, at the individual and at the aggregate level of granularity. Privacy preservation may be linked to anonymization techniques that are required to prior to disclosure of customer data, such as: k-anonymity, pseudonymization, and redaction of Personally Identifiable Information (PII). Recent research, in the areas of differential privacy techniques, offers mathematically proven privacy-preservation techniques. Differential privacy seeks to preserve privacy (minimize risk) while maintaining data utility. Differential privacy does not guarantee perfect privacy or perfect utility. While differential privacy is not perfect, it is generally more acceptable than PII redaction or pseudonymization when re-identification of a single individual is of concern.

*G. Industrial Big Data Analytics Paradigms: Batch vs. Realtime Streaming*

Big data analytics in manufacturing spaces is the process of using machine learning algorithms running on powerful computing platforms to uncover potentials concealed in huge manufacturing data region, such as hidden patterns or unknown correlations. According to the processing timeliness requirement, industrial big data analytics can be categorized into two alternative paradigms(see Fig.2):

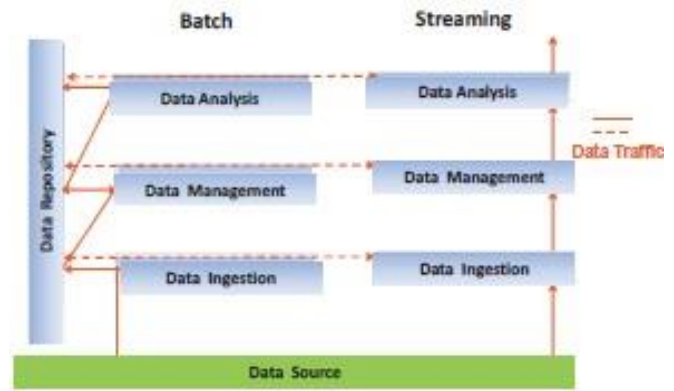

Fig. 5. Industrial big data analytics processing modes.

*1) Batch Processing:* In the batch-processing paradigm, Data from different sources are first stored into a big data analytics platform and then analyzed. This model of processing is well suited for applications that need a large amount of data to produce accurate, meaningful intelligence. Typically time is not a pressing factor in consideration. Moreover, the specific type of analytical operations may not be fully understood nor anticipated at collection time. The batch processing model provide tradeoffs between data availability and data storage size limits. The more data available for later processing, the more storage it needs to keep these data. It also provides tradeoffs between analytics accuracy and reporting time. The more through the analytics computation, the longer it takes to get the results. MapReduce has become the dominant batch processing model. The core idea of MapReduce is that data are first divided into small chunks. Next, these chunks are processed in parallel and in a distributed manner to generate intermediate results. The final result is derived by aggregating all the intermediate results. This model schedules computation resources close to data location, which avoids the communication overhead of data transmission. Transfer of data between the different layers is represented below.

*2) Streaming processing:* In real-time stream processing solution paradigm, data arrives in a stream. In its continuous arrival, because the stream is fast and carries enormous volume, only a small portion of the stream is stored in limited memory. This characterizes data which is received in online/real-time mode by the Industrial Big Data Analytics platform for immediate processing. The focus of this model is to provide accurate analytical results based on current data available. The data may or may not be stored for future use. Transfer of data between the different layers is represented below. Streaming processing theory and technology have been studied for decades. Representative open source systems include Spark, Storm, S4,and Kafka.

IV. APPLICATION IN INDUSTRIAL BIG DATA

*A. Smart Factory Visibility*

Based on manufacturing companies implemented sensors and computerized automation for decades, the sensors, Programmable Logic Controllers (PLC) and PC-based controllers and management systems are largely connected from IT and online operational systems by Industrial IoT technologies. These Factory systems are organized in hierarchical fashion within individual data silos and often connections to internal systems. Our Big data analytics Platform will provide production line information to decision makers and improve factory efficiency. Fig6 shows that GE Smart factory visibility solution allow devices to display performance data and status updates.

Instead of being chained to a control room, facilities managers and production personnel will have easy access to real-time information and collaborate more effectively. Mark Bernardo, the general manager of automation software for GE Intelligent Platforms, says " When you equip people with mobile technology, you can dramatically shrink the delta between when a problem occurs and when its acted upon. If theres a quality control problem in a production line, they can shut down the line before it continues to create products that will all be waste.

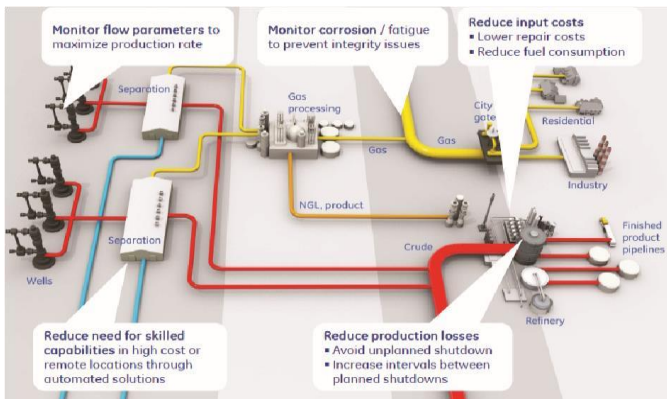

Fig. 6. GE Smart factory visibility solution from the oil and industry.

The benefits of big data analytics visibility will extend beyond the enterprise to a wide range of suppliers and third party providers of services, production and capital goods. Industrial Big data analytics will enable extensive involvement by third party suppliers in the direct operations and maintenance of manufacturing plants with new service and supply business models based on increased visibility and remote monitoring. Suppliers of capital equipment may now be able to offer business models that involve production based revenue rather than capital equipment sales if equipment can be monitored sufficiently for both output and maintenance status. Parts, services and production suppliers within Maintenance, Repair and Operations (MRO) will use big data analytics to monitor distributed inventories, tank levels of process fluids, wear parts conditions, and production rates. This will create entirely new and very closely linked business relationships between manufacturers and their suppliers.

*B. Machine Fleet in Industrial Big Data Environment*

Machine fleet is very common that identical machines are being exposed to completely different working conditions for different tasks. In contrast, most predictive and prognostic methods are designed to support a single or limited number of machines and working conditions. Currently, available prognostic and health management methods are not taking advantage of considering these identical machines as a fleet by gathering worthwhile knowledge from different instances. At the same time, large-scale different instances data have been isolated from each other and from local and distant business networks. Today, we use our Big data analytics Platform to integrate every instance within a plant and provide connectivity and information sharing across multiple locations and business processes. Once machinery and systems are connected within the plant, manufacturers can use this information to automate workflows to maintain and optimize production systems without human intervention. One example of this is machine fleet in industrial big data environment(see Fig.7). The company installed software that keeps a record of how different equipment is performing, such as the speed of fans in the painting booth. The software can automatically adjust the machinery if it detects that a measurement such as fan speed, temperature, or humidity has deviated from acceptable ranges.

*C. Energy Management*

In many industries, energy is frequently the second largest operating cost. But many companies lack cost effective measurement systems and modeling tools or performance and management tools to optimize energy use in individual production operations, much less in real-time across multiple operations, facilities, or an entire supply chain. There are numerous ways that IoT and automation of environmental controls such as HVAC and electricity can create cost savings for manufacturers. Base on our Big Data Analytics Platform, building whole energy solutions can provide peak demand charge avoidance and enable economy model operations. IoTenabled Energy Management also offer integrated weather data and prediction analysis to help manufacturers understand expenses and plan energy usage.

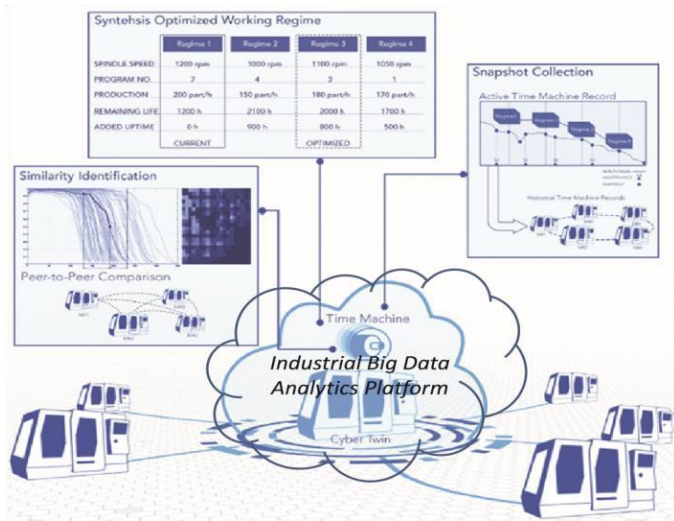

Fig. 7. Machine fleet in industrial big data environment [49].

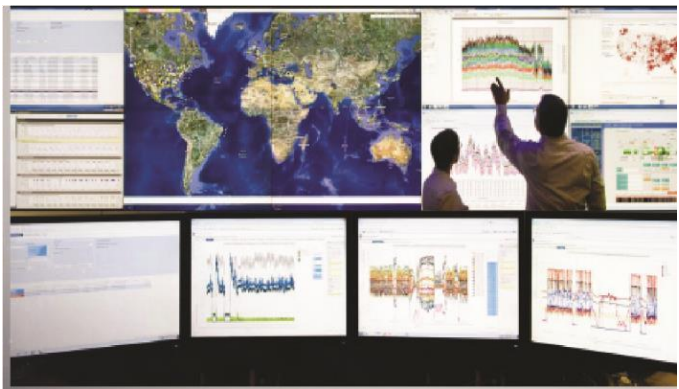

Fig. 8. IoT-enabled energy management system from the oil and industry.

*D. Proactive Maintenance*

Manufacturers have widely accepted the concept of preventative and condition-based monitoring but many are still in the process of implementing these programs. Lower cost sensors, wireless connectivity and big data processing tools make it cheaper and easier to collect actual performance data and monitor equipment health. The big data analytics platform means being able to assess the current or past condition of each machine, and react to the assessment output. Such health assessment in big data environment can be performed by using a data-driven algorithm to analyze data/information collected from the given machine and its ambient environment. The condition of the real-time machine can be fed back to the machine controller for adaptive control and machine managers for in-time maintenance.

If the manufacturer has equipment which is supposed to operate within whole manufacturing process, industrial big data analytics platform can extract meaningful information from big data more efficiently, and further perform more intelligent decision-making. The main objective of design, control and decision-making of machine operations is to meet the production goal with effective and efficient production planning and maintenance scheduling. The actual system performance often deviates from the designed productivity target because of low operational efficiency, mainly due to significant downtime and frequent machine failures. In order to improve the system performance, two key factors need to be considered: (1) the mitigation of production uncertainties to reduce unscheduled downtime and increase operational efficiency, and (2) the efficient utilization of the finite resources on the critical sections of the system by detecting its bottleneck components. With the advent of PHM in a CPS framework, rich PHM knowledge is utilized to assist and enhance the capability of decision-making in production control and maintenance scheduling to achieve high reliability and availability.

*E. Just in Time Supply Chain*

Just in time manufacturing has been important concept for Smart Supply Chain. Our Big data analytics Platform can help manufacturers gain a better acquisition of the supply chain information that can be delivered in realtime. By integrating the production line and balance of plant equipment to suppliers, all parties can understand interdependencies, the flow of materials, and manufacturing cycle times. Our Big data analytics can archieve for location tracking, remote health monitoring of inventory, and reporting of parts and products as they move through the supply chain, among many other things. Our Big data analytics can also collect and feed delivery information into an ERP system; providing up-todate information to accounting functions for billing. Realtime information process in our Big data analytics will help manufacturers identify issues before they happen, lower their inventory costs and potentially reduce capital requirements.

The complex event processing realize real-time monitoring of almost every link of the supply chain, ranging from commodity design, raw material purchasing, production, transportation storage, distribution and sale of semi-products and products. It is also possible to obtain products related information, promptly, timely, and accurately so that enterprises or even the whole supply chain can respond to intricate and changeable markets in the shortest time. The reaction time of traditional enterprises is 120 days from requirements of customers to the supply of commodity while advanced companies that make use of these technologies (such as Walmart and Metro) only take few days allowing them to work with zero safety stock. Additionally, complex event processing technology can real-time interact to the ERP program helps the shop assistants to better inform customers about availability of products and give them more product information in general.

V. CONCLUSION

With the rapid advancement of Information and Communication Technologies (ICT) and the integration of advanced analytics into devices, manufacturing, products and services, manufacturers are facing new challenges of maintaining their competency, increasing production efficiency, improving production quality, reducing energy consumption, and reducing cost. Big data analytics become the foundation for manufacturing area such as forecasting, proactive maintenance and automation. In this survey, we have focused on five key techniques for designing and implementing efficient and high performance industrial big data analytics platform, relating to: (1) highly distributed industrial data ingestion, (2) industrial big data repository, (3) large-scale data management, (4) industrial big data analytics, (5) industrial big data governance. We reviewed the industrial big data analytics and some advanced technologies such as Apache Hadoop, GE big data plaform and SAP Hana, which provide the basis for industrial big data analytics. In addition, we also provided the five typical applied case of industrial big data, which including smart factory visibility, machine fleet, energy management, proactive maintenance, and just in time supply chain. These newly methodologies and typical applied case will not only help manufacturers make more relevant and valuable decisions, but also provide capability of predictive manufacturing and service innovations. This survey also provides a comprehensive review of important technology in industrial big data analytics of related works to date, which hopefully will be a useful resource for further industrial big data analytics research.

ACKNOWLEDGMENT

The authors also would like to thank anonymous editor and reviewers who gave valuable suggestion that has helped to improve the quality of the manuscript. This research has been supported This work was supported in part by the National Key R&D Program of China (No.2016QY03D0501); by the National Natural Science Foundation of China under Grant 61772525, Grant 61772524, Grant 61702517, Grant 61602482; by the Beijing Municipal Natural Science Foundation under Grant 4182067.